# 2D polyphthalocyanines (PPCs) of different structure and polymerization degree: chemical factors, characterization and processability

Daria M. Sedlovets, Vladimir T. Volkov, Igor I. Khodos, Alexandr V. Zotov, Vitaly I. Korepanov*

Institute of microelectronics technology and high purity materials RAS

* *e-mail*: korepanov@iptm.ru

**Abstract**

2D conjugated polyphthalocyanines can be obtained as two distinctly different types of material with specific molecular structures and different morphological properties. It was believed that the temperature is the key factor affecting the chemical reaction, but we show that even at the optimal temperature (420°C), the reaction on vapor/solid interface and liquid/solid interface yields different products: while the former is well-polymerized and ordered, the latter is amorphous and cross-linked with the typical conjugation scale of single PC ring. IR spectroscopy is most sensitive tool for identifying the molecular structure, providing the information on polymerization degree, structural uniformity and content of terminal groups. We show that, unlike the ordered PPCs, the cross-linked product can be reproducibly obtained as continuous conductive material.

**Introduction**

2D conjugated polyphthalocyanines (PPCs) are actively studied as promising materials for next generation batteries[1], advanced fuel cells[2], water splitting catalysts[3], magnetic storage devices[4], spintronics[5] and other applications. Despite the long study history of PPCs[6,7], there is a controversy in the literature regarding the structure and characterization of these materials. Several groups reported significantly different spectra of the material, produced by different techniques[1,8–12]. It has been shown recently that the IR, Raman and UV-vis spectra of PPCs strongly depend on the polymerization degree, and products, obtained in many preceding works, should be assigned to either monomeric polyphthalocyanines or to material of low polymerization degree (oligomers)[13]. A general way to produce PPCs is reaction of pyromellitic tetranitrile with metals or salts at elevated temperature (fig. 1), and it was believed that the temperature is the key factor influencing the reaction pathway: the formation of polymer needs >350°C, while at ~200°C only single ring can be formed (octacyano phthalocyanine, OCP) [11,12,14–17]. In the present work, we show that there are other chemical factors strongly affecting the structure and polymerization degree of PPCs. Even at the optimal reaction temperature (~420°C) two distinctly different products can be obtained depending on the chemical set-up.

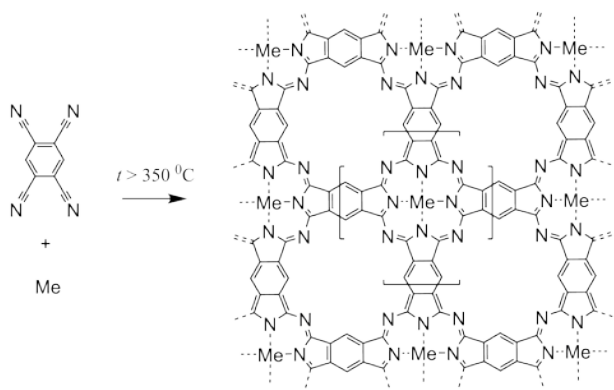

Figure 1. General synthetic route to PPCs: reaction of PMTN with metal source (Me=Fe, Co, Ni… or $H_2$)

If we use elementary metals (not salts) as a metal source, then the reaction is heterogeneous. We can perform it in two modes: as a vapor/solid and liquid/solid set-up. PMTN has a melting point of ~266°C, therefore at the typical reaction temperatures it should be in a liquid state. Importantly, when heated in atmosphere, PMTN changes color to blue above 200°C, which indicates the formation of OCP, a first step of the polymerization[14].

We study the reaction of PMTN with a series of transition metals (Cr, Fe, Co, Ni, Cu) as well as metal-free ($H_2$) and show that the polymerization degree depends on the chemical set-up rather than on the metal.

For a number of microelectronics-related applications, it is desirable to obtain the material in the form of continuous conductive film. This can be difficult, because such conventional processing techniques as spin-coating, sputtering or thermal evaporation cannot be applied to PPCs[9]. We show that morphology and conductivity of PPC films are also influenced by the production method.

**IR and Raman spectra**

From the IR spectra, we can see significantly different products from the reaction at liquid/solid and at vapor/solid interfaces at the same temperature (at 420°C), fig. 2. While the spectrum after the liquid/solid regime resembles the broadened spectrum of the monomeric PCs, the product from the vapor/solid reaction shows small number of relatively narrow bands at different positions. It was proved recently with the help of DFT calculations that the 'true' spectrum of well-ordered polymer is the latter, while the former is poorly-polymerized (cross-linked) product, the electronic conjugation in which has a characteristic scale of single PC ring.

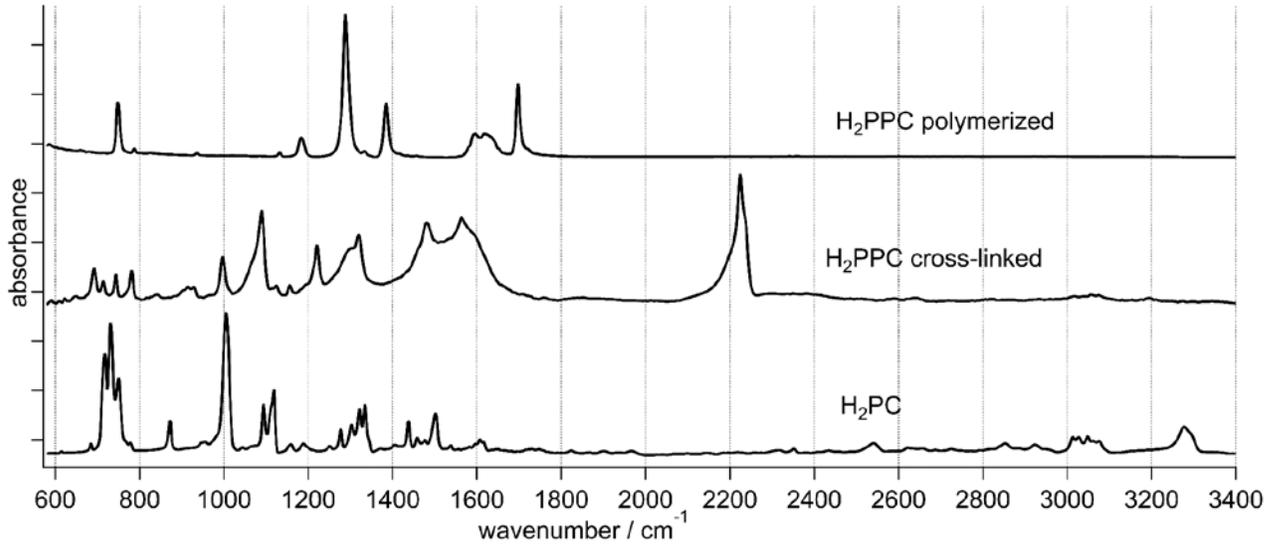

Figure 2. IR spectra of well-polymerized $H_2PPC$ from the synthesis at vapor/solid interface (top), cross-linked product from the liquid/solid interface (middle) and the spectrum of the "monomeric" $H_2PC$ (bottom).

Contrary to intuition, well-polymerized PPCs lack the characteristic spectral features of monomeric phthalocyanines. Moreover, the spectra of the polymers have significantly less bands than those of the monomeric PC. This is the consequence of the fact that while the PC has 57 atoms in the molecule, the translational unit in the polymer has only 33 atoms (fig. 1). In the $D_{4h}$ point group (symmetry of PC molecule and of perfect sheet of 2D polymer), there are total 36 symmetry-allowed bands in the IR spectra of monomer as compared to 22 in that of the polymer. Not all of these bands have high intensity, and not all of them fall in the fingerprint region, which makes the IR spectrum of PPC quite simple[13].

Similarity between the spectra of cross-linked PPCs and monomeric PCs is an evidence of low polymerization degree. Another important feature is high intensity of terminal groups (-C≡N) at ~2220 $cm^{-1}$ in the spectra of cross-linked material as compared to the ordered one (fig. 2). In terms of structure uniformity, this means a high number of lattice defects in the former. The bandwidths of cross-linked and ordered polymers reflect well this structure difference: the bands of the ordered material are much narrower (characteristic FWHM is about 15 cm-1).

Raman spectra of PPCs are significantly different from those of the monomeric PCs (fig. 3). This difference results from the electronic structure: the 2D conjugated π-electronic system of PPCs resembles that of patterned graphene; electronic conjugation is likely to create a resonant Raman effect. The cross-liked product shows intermediate spectral pattern between the polymerized one and the monomer. The resonance however is likely to enhance the cross-section of more conjugated structure fragments, therefore the similarity between the conjugated and cross-liked products is more pronounced in Raman spectra then in the IR. DFT calculations confirm that the well-polymerized PPCs should give several intense close-lying bands in higher fingerprint region[13].

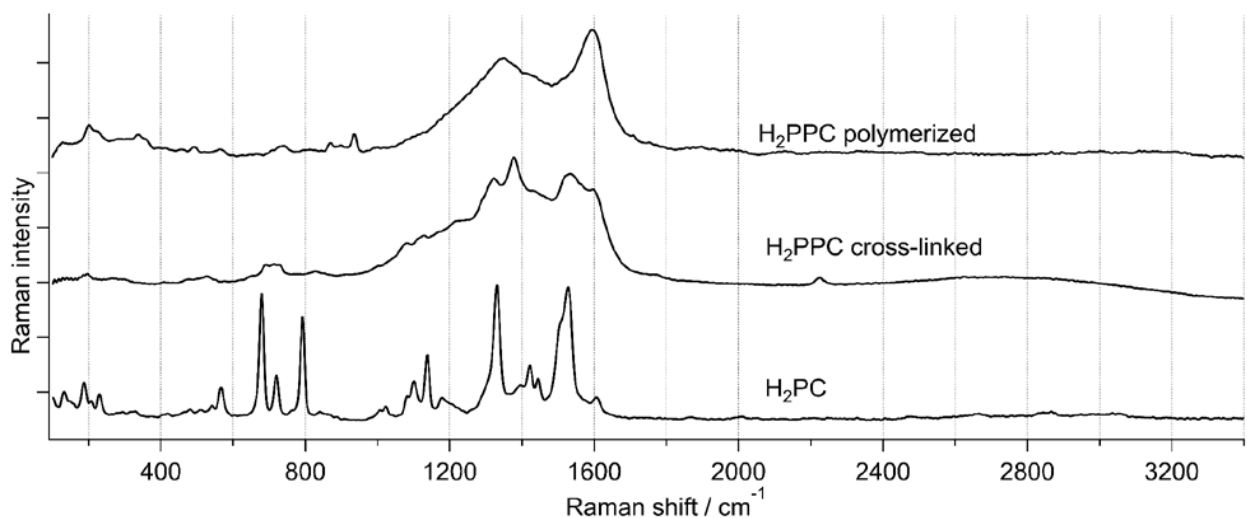

Figure 3. Raman spectra of well-polymerized $H_2PPC$ from the synthesis at vapor/solid interface (top), cross-linked product from the liquid/solid interface (middle) and the spectrum of the "monomeric" $H_2PC$ (bottom).

Figs. 2 and 3 shows IR and Raman spectra of $H_2PPC$ from liquid/solid and vapor/solid reaction modes as compared to the spectra of the monomeric $H_2PC$. The other studied MePPCs (Me= Cr, Fe, Co, Ni) show analogous spectral patterns in the two set-ups (SI section 1 and 2). In the fingerprint region, the bands in MePC spectra belong to the vibrations of organic skeleton of the molecule, therefore show only minor difference upon changing the central ion[18]. The spectra of CuPPC are reported elsewhere[9,12], and they follow the same tendency.

**TEM**

IR spectra show pronounced difference in molecular structure between ordered and cross-linked materials. The corresponding difference at nanoscale is also observed in TEM images (fig. 4): the polymerized product shows ordered (crystalline) domains with the typical size of few tens of nm, while the cross-linked product has an amorphous structure.

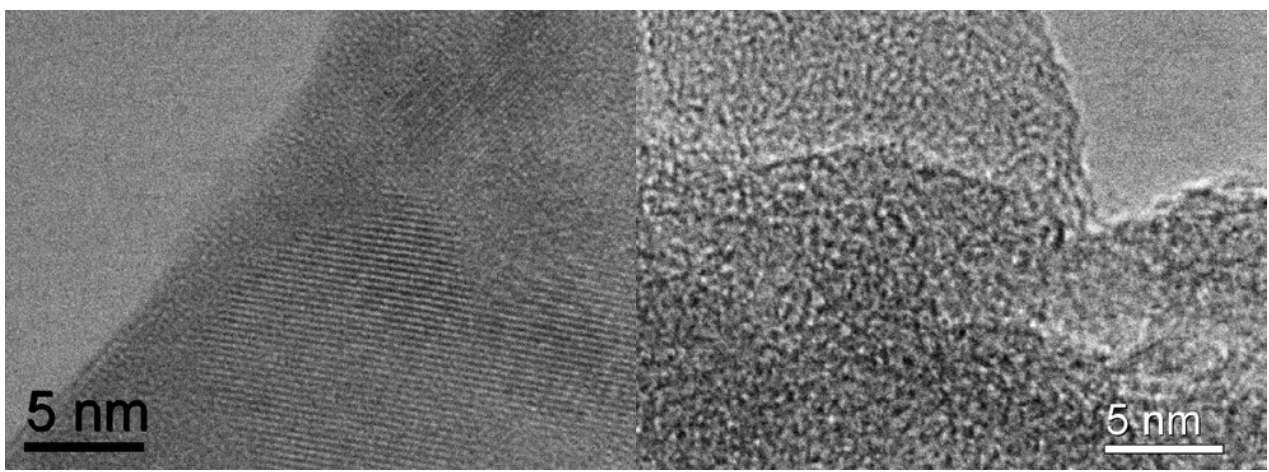

Figure 4. TEM images of ordered (left) and cross-linked (right) H$_2$PPC materials

**Continuity, conductivity and the growth mechanism**

One unexpected finding was that the continuity and conductivity of ordered and cross-linked products was significantly different. The ordered polymer did not always give the conducting film, while the cross-linked material showed well reproducible conductivity. Together with the structural information, this fact provides understanding of the growth mechanism in the two regimes. Thin films from liquid/solid mode take the morphology of the PMTN layer as it melts and polymerizes. This makes it continuous and conductive. On the contrary, in the vapor/solid mode the growth proceeds along the crystal facets and/or catalyst particles, which does not always yield the continuous film. Another important factor that affects the morphology is the formation of the islands and loss of continuity of the few-nm metal film upon heating.

Quantum-chemical calculations suggest that MePPCs are low-gap semiconductors with HOCO-LUCO energy difference of 0.02 eV[13]. It was shown that both HOCO and LUCO originate from the π-electrons of the organic skeleton, therefore all MePPCs should have similar electronic properties. In the present work, we performed the electrical measurements for H$_2$PPC and CrPPC. The characteristic values of resistances at room temperature are 3.6 GΩ/square and 1.4 MΩ/square for H$_2$PPC and CrPPC, respectively. For CrPPC, the nearly-linear resistance decrease was observed with increasing temperature. We may conclude that the conductivity has the behaviour expected for low-gap semiconductor. Details on the electric measurements are provided in the supplementary information (SI section 3). PPCs show a big promise to a variety of microelectronics applications [1,2,4,5], therefore their electrophysical properties require further study.

**Experiment**

The vapor/solid synthesis of MePPCs was done according to the previously reported procedure[8,9]. Briefly, a thin layer (few nm) of metal was deposited on a substrate and placed in a reactor for chemical vapor deposition (CVD). Therein, it was exposed to the PMTN vapor at 420°C for a few hours. As a substrate, KBr plate or silicon wafer was used. We should emphasize that the vapor/solid reaction is selective even for the metal-free PPC, i.e. the polymer grows only on the catalyst, while the reactor walls remain practically clear.

Deposition of few-nm metal films on the substrate was done with electron-beam evaporation system (Leybold L560), the procedure details are described elsewhere [19]. For $H_2PPC$, instead of metal, we used pre-coating with molybdate catalyst. An aqueous solution (1 g/l) of ammonium heptamolybdate (99% mass $(NH_4)_6Mo_7O_{24}*4H_2O$ by Reachem) was drop-casted on a KBr substrate (about 0.05 ml per 5x5 mm wafer) and then dried at 150°C.

For the liquid/solid regime, the same CVD set-up was used. The difference was that the sample with a portion of PMTN was wrapped in aluminum foil instead of exposing to the PMTN vapor; the pressure of hydrogen was maintained at ~1atm.

In the synthesis, hydrogen plays several roles: remove oxygen, suppress surface oxides and, in case of liquid/solid regime, to maintain the pressure. Both in our experiments and in the literature[1], the polymerization products show the ~1780 $cm^{-1}$ carbonyl band, unless the oxygen is completely excluded from the reactor.

FTIR spectra were taken with a Bruker Vertex 70V spectrometer combined with a HYPERION 2000 IR-microscope in the 400-4000 $cm^{-1}$ range under 1 $cm^{-1}$ resolution. Raman spectra were measured with a Bruker Senterra micro-Raman system under 532 nm excitation and laser power 2 mW. TEM images were acquired with a JEOL JEM 2000FX transmission electron microscope.

Electrical measurements were carried out by SourceMeter Keithley SMU 2450 in RTI optCRYO 105 cryostat cooled with liquid nitrogen. The temperature was controlled by multimeter system Keithley DMM 2701.

**Conclusions**

Temperature was believed to be the key factor controlling the synthesis of 2D polyphthalocyanines. In this work we showed however that even at the optimal temperature, the reaction can be performed in two different regimes: heterogeneous (vapor/solid interface) and quasi-homogeneous (liquid/solid interface). These two modes yield two kinds of polymers: ordered and cross-linked, correspondingly. The two products can be easily distinguished by the IR spectra, in which the cross-linked polymer shows spectral features resembling the monomeric PCs with intense signal from terminal groups (-C≡N), while the ordered one has less number of bands with narrower bandwidth and lower intensity of terminal groups. Raman spectral patterns of PPCs are also significantly different from those of the monomers due to extension of 2D electronic conjugation.

For the well- ordered PPCs, the growth proceeds along crystal facets and/or catalyst particles, which does not always yield conductive films. On the contrary, the cross-linked PPC films are continuous and show well reproducible conductivity. The material has an electric behaviour of low-gap semiconductor.

**Acknowledgements**

This work was supported by Russian Science Foundation (project no. 17-73-10128).

**References**

[1] Chen J, Zou K, Ding P, Deng J, Zha C, Hu Y, Zhao X, Wu J, Fan J and Li Y 2018 Conjugated Cobalt Polyphthalocyanine as the Elastic and Reprocessable Catalyst for Flexible Li-CO2 Batteries *Adv. Mater.* 1805484

[2] Zhang Z, Dou M, Liu H, Dai L and Wang F 2016 A Facile Route to Bimetal and Nitrogen-Codoped 3D Porous Graphitic Carbon Networks for Efficient Oxygen Reduction *Small* **12** 4193–9

[3] Zhang Z, Qin Y, Dou M, Ji J and Wang F 2016 One-step conversion from Ni/Fe polyphthalocyanine to N-doped carbon supported Ni-Fe nanoparticles for highly efficient water splitting *Nano Energy* **30** 426–33

[4] Wang P, Jiang X, Hu J, Huang X and Zhao J 2016 Giant magnetic anisotropy of a 5d transition metal decorated two-dimensional polyphthalocyanine framework *J. Mater. Chem. C*

[5] Zhou J and Sun Q 2011 Magnetism of Phthalocyanine-Based Organometallic Single Porous Sheet *J. Am. Chem. Soc.* **133** 15113–9

[6] Drinkard W C and Bailar J C 1959 Copper Phthalocyanine Polymers *J. Am. Chem. Soc.* **81** 4795–7

[7] Epstein A and Wildi B S 1960 Electrical Properties of Poly-Copper Phthalocyanine *J. Chem. Phys.* **32** 324–9

[8] Sedlovets D M, Shuvalov M V., Khodos I I, Trofimov O V. and Korepanov V I 2018

Synthetic approach to thin films of metal-free polyphthalocyanine *Mater. Res. Express* **5** 026401

[9]     Sedlovets D M, Shuvalov M V., Vishnevskiy Y V., Volkov V T, Khodos I I, Trofimov O V. and Korepanov V I 2013 Synthesis and structure of high-quality films of copper polyphthalocyanine – 2D conductive polymer *Mater. Res. Bull.* **48** 3955–60

[10]    Abe K, Ohkatsu Y and Kusano T 1988 Preparation of metal-free polyphthalocyanine thin films by an evaporation-polymerization method *Die Makromol. Chemie* **189** 761–4

[11]    Wöhrle D, Schmidt V, Schumann B, Yamada A and Shigehara K 1987 Polymeric Phthalocyanines and their Precursors, 13. Synthesis, Structure and Electrochemical Properties of Thin Films of Polymeric Phthalocyanines from Tetracarbonitriles *Ber. Bunsenges. Phys. Chem.* **91** 975–81

[12]    Yudasaka M, Nakanishi K, Hara T, Tanaka M, Kurita S and Kawai M 1987 Metal phthalocyanine polymer film formation by the double source evaporation of tetracyanobenzene and metal *Synth. Met.* **19** 775–80

[13]    Korepanov V I and Sedlovets D M 2019 Spectroscopic identification of conjugated polymeric phthalocyanines *Mater. Res. Express* **accepted** ArXiv preprint 1812.00708

[14]    Koudia M and Abel M 2014 Step-by-step on-surface synthesis: from manganese phthalocyanines to their polymeric form. *Chem. Commun. (Camb).* **50** 8565–7

[15]    Wöhrle D 2001 Phthalocyanines in Macromolecular Phases – Methods of Synthesis and Properties of the Materials *Macromol. Rapid Commun.* **22** 68–97

[16]    Wöhrle D, Marose U and Knoop R 1985 Polymeric phthalocyanines and their precursors, 8. Synthesis and analytical characterization of polymers from 1,2,4,5-benzenetetracarbonitrile *Die Makromol. Chemie* **186** 2209–28

[17]    Nardi E, Koudia M, Kezilebieke S, Bucher J-P and Abel M 2016 On-Surface Synthesis of Phthalocyanine Compounds *On-Surface Synthesis* Advances in Atom and Single Molecule Machines ed A Gourdon (Cham: Springer International Publishing) pp 115–29

[18]    Liu Z, Zhang X, Zhang Y and Jiang J 2007 Theoretical investigation of the molecular, electronic structures and vibrational spectra of a series of first transition metal phthalocyanines *Spectrochim. Acta, Part A Mol. Biomol. Spectrosc.* **67** 1232–46

[19]    Volkov V, Levashov V, Matveev V, Matveeva L, Khodos I and Kasumov Y 2011 Extraordinary Hall effect in nanoscale nickel films *Thin Solid Films* **519** 4329–33

# Supplementary Information

2D polyphthalocyanines (PPCs) of different structure and polymerization degree: chemical factors, characterization and processability

Daria M. Sedlovets, Vladimir T. Volkov, Igor I. Khodos, Alexandr V. Zotov, Vitaly I. Korepanov

## 1. IR spectra of a series of MePPCs

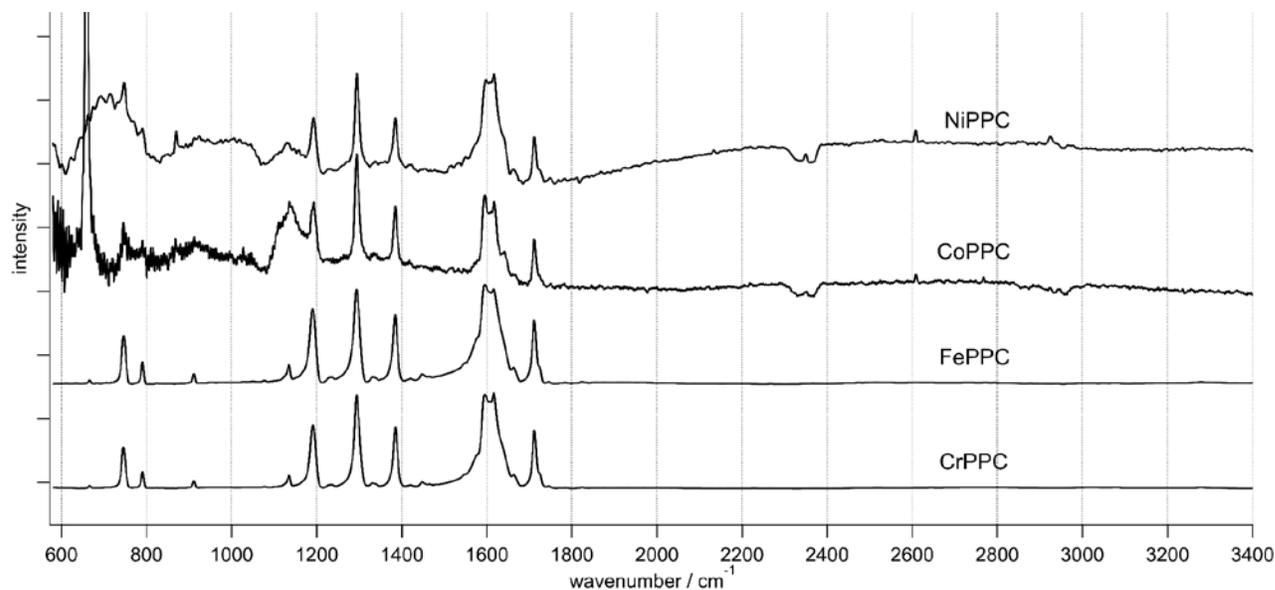

Figure 1. IR spectra of MePPCs from vapor/solid regime

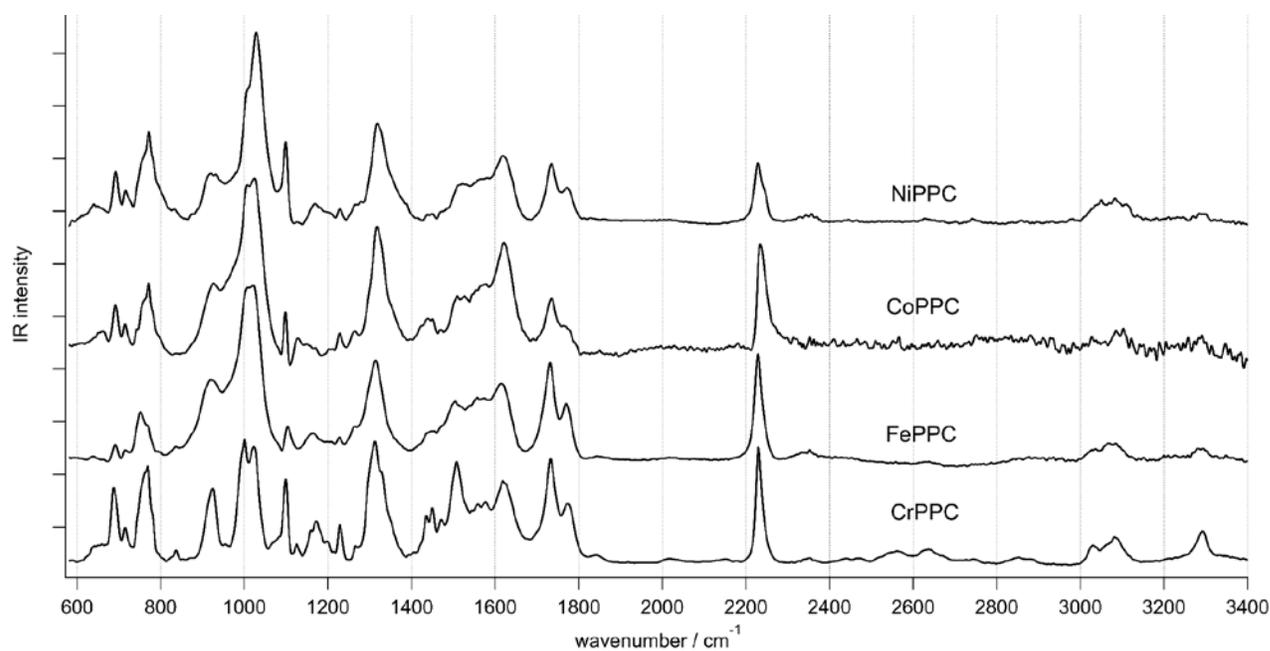

Figure 2. IR spectra of MePPCs from liquid/solid regime

## 2. Raman spectra of a series of MePPCs

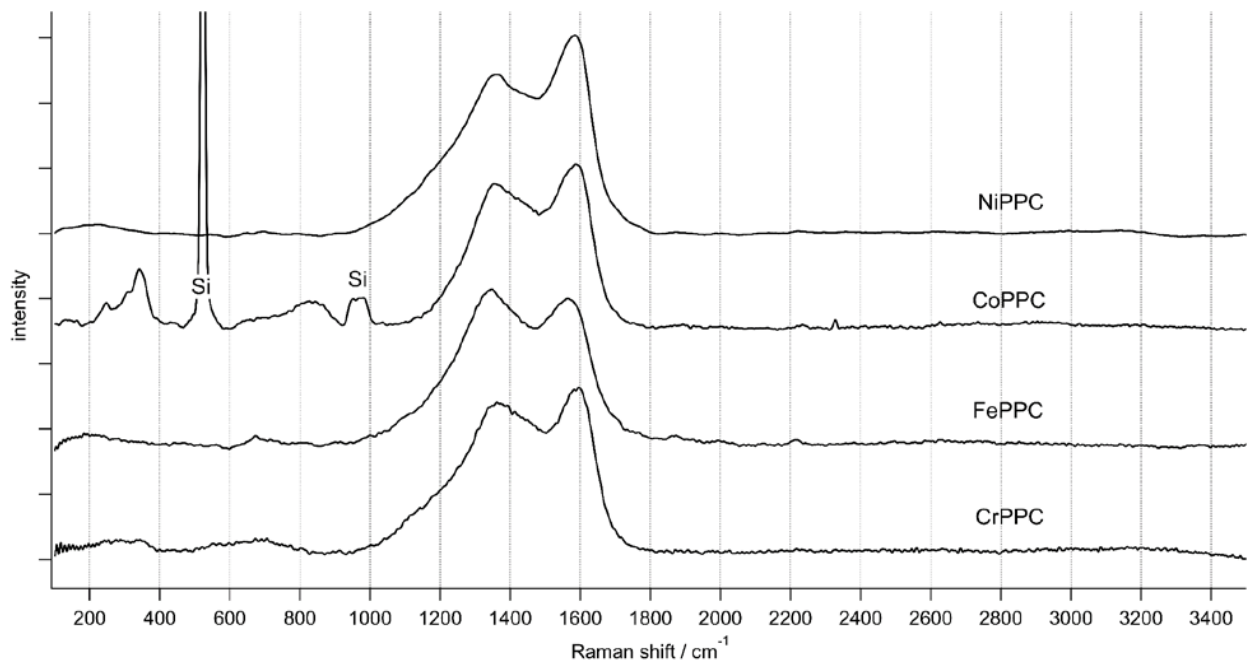

Figure 3. Raman spectra of MePPCs from vapor/solid regime

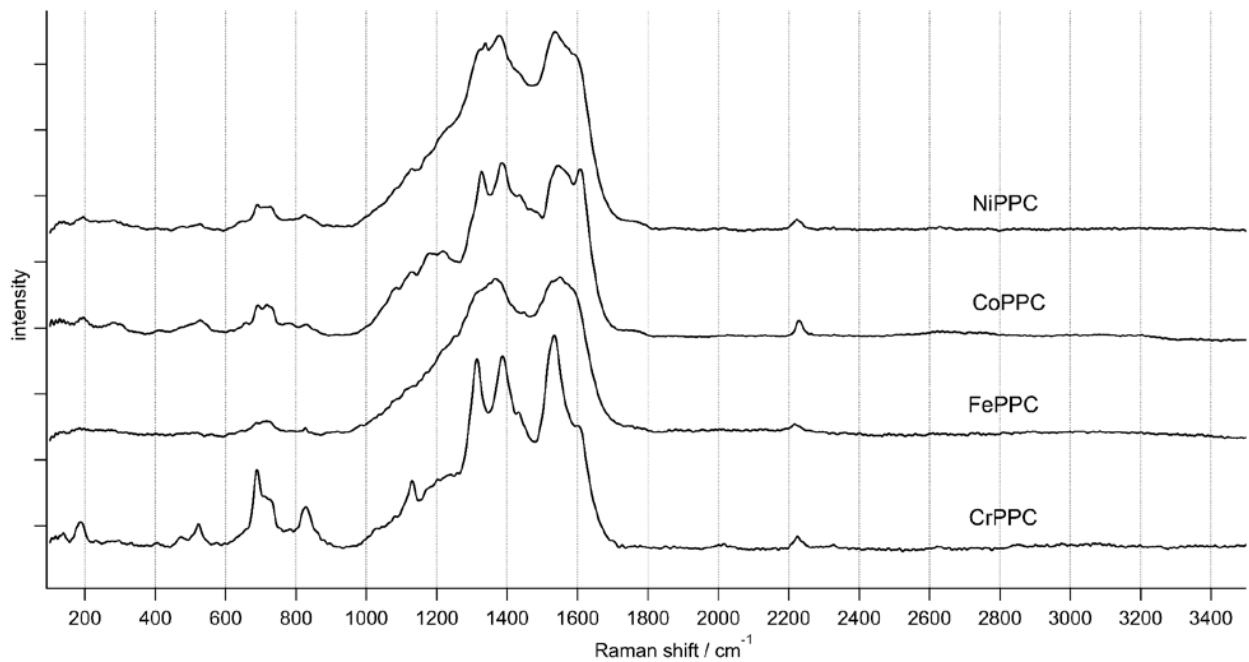

Figure 4. Raman spectra of MePPCs from liquid/solid regime

## 3. Electric measurements

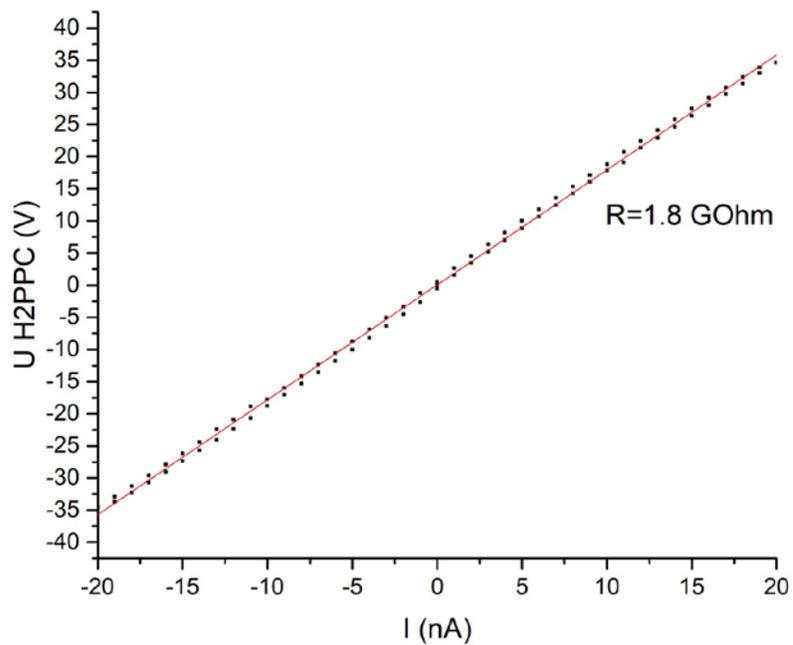

Figure 5. U(I) for H$_2$PPC @ T=293*K.

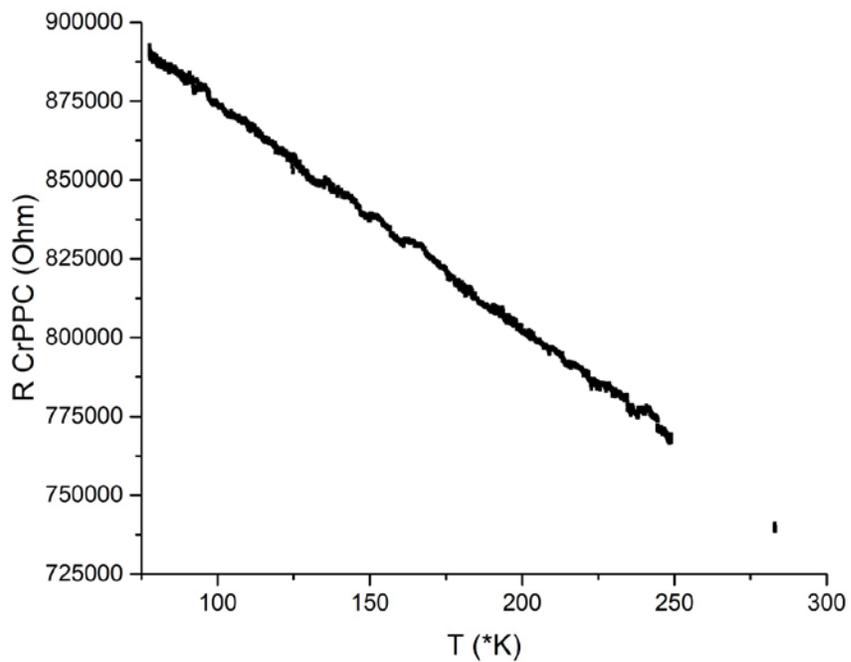

Figure 6. R(T) for CrPPC. Unusual linear temperature dependence.